\documentclass{birkjour_t2}

\usepackage{color}

\newcommand{\xdif}{\mathrm{d}}
\newcommand{\xDif}{\mathrm{D}}

\begin{document}

\title{Formulation of small-strain magneto-elastic problems}

\author{Tuomas Kovanen, Timo Tarhasaari and Lauri Kettunen}
\address{Electromagnetics, Department of Electrical Engineering, Tampere University of Technology, Tampere, Finland.}
\email{tuomas.kovanen@tut.fi, timo.tarhasaari@tut.fi, lauri.kettunen@tut.fi}

\begin{abstract}Despite of the topical engineering need and all scientific investments, the mathematical formulation of modeling elastic deformations in magnetic systems is not yet fully established. Often, especially in electrical engineering applications, a model assuming small (infinitesimal) strains seems sufficient. To express such small-strain magneto-elastic problems in a suitable form for discretization methods, we present here a formulation in the framework of differential geometry. The given analysis shows algebraic similarity between small-strain elasticity and magnetism. This suggests that a class of magnetic, elastic, and magneto-elastic problems may be modeled in the same algebraic category, constituting suitable domain for discretizations.
\end{abstract}



\maketitle

\section{Introduction} \label{intro}

A class of electromechanical problems couple magnetism and elasticity. A magneto-elastic coupled model is relevant for the development of electric motor, sensor and actuator technologies, and much work have been done to make such modeling available to engineers \cite{Besbes_96,Besbes,Bossavit_92,Bossavit_07,Brown,Dorf,Fonteyn,Liu,Mini,Ogden,Ren_95,Stein}. In modeling, it is commonplace to rely on a variational formulation of the coupled problem. Variational formulation is also adopted in a recent paper by Bossavit \cite{Bossavit}. Therein, a small-strain magneto-elastic model is built on an analogy between magnetism and small-strain elasticity, with emphasis on finite element implementation using Whitney elements. 

In the present paper we consider the formulation of small-strain magneto-elastic problems using analysis on manifolds. Accordingly, we will be precise about the mathematical structures used in different parts of the analysis. This is how mathematics deals with the same thing which is called coordinate system invariances in physics. For example, defining Maxwell's equations on differentiable manifold means the corresponding coordinate equations are invariant to general coordinate transformations. On the other hand, to express certain equations in a coordinate independent manner, we will introduce additional structure on the manifold. Example of an equation which we express by using additional structure --- that of a connection --- is the mechanical equilibrium equation. In short, the manifold theory serves to drive information from coordinates to well defined mathematical structures. This makes possible a condensed presentation, allowing rapid access to the magneto-elasticity problem for mathematically informed reader.

As a prerequisite we assume that the reader is familiar with differential forms, and the basic tools of differential geometry. In particular, one recognizes the exterior derivative as the counterpart for the gradient, curl and divergence of vector analysis. For example, the Gauss's law and Ampere's law in magnetostatics are naturally expressed by using the exterior derivative, whose definition requires only the differentiable structure. On the other hand, the magnetic constitutive law requires more structure, and is expressed in the variational formulation by using a magnetic energy density object.

For algebraic similarity between elasticity and magnetism, we will introduce the reader to vector- and covector-valued differential forms together with their natural product \cite{AMR,Frankel,Kanso,Kovanen,Segev,Yavari}. These notions require only the differentiable structure. In addition, we present some related tools of differential geometry. Among these there is the covariant exterior derivative, whose definition requires the additional structure of a connection \cite{Kanso,Kovanen,Yavari}. Similarly to magnetism the constitutive law between stress and strain will be expressed in the variational framework by using an energy density object.

When it comes to magneto-elasticity, a variational formulation involves, in general, magneto-elastic energy density. Such magneto-elastic energy density constitutes a way to express ``coupled constitutive laws'', and it implies that magneto-elasticity cannot be meaningfully separated into ``magnetic'' and ``elastic'' parts. Such genuine magneto-elastic coupling is often called magnetostriction \cite{Billard,Bossavit}. Here, we give two alternative mathematical constructs for magneto-elasticity, both of them allowing magnetostriction.

The given analysis shows that certain problems of magnetism, small-strain elasticity, and small-strain magneto-elasticity may be modeled by using geometric constructs that are in many ways similar algebraically. This makes it relevant to search for \emph{algebraic category} for the modeling of a class of such problems. We argue that the detailed specification of an appropriate category should be made such that (i) it is general enough to cover important problems, and (ii) its discretization to the category of finite dimensional vector spaces can be realized. An analysis such as the one given here is a necessary step in the way to specify an appropriate category.

\section{Space-time, material bodies, and allowable observers} \label{setting}

To position the formulation of this paper into a more general context, we begin by introducing material bodies, the underlying (classical) space-time, and the notion of observer. These notions are taken from Ref. \cite{Marsden}, and are reproduced here (using slightly different notation) to make the paper self-contained. Moreover, we specify here the observers allowed by the given formulation.

In classical space-time, the notion of simultaneity has absolute meaning, and simultaneous events may be identified with a reference manifold. By observation of space-time we mean the process of making such an identification for all time instants. Independently of observation, it is assumed that time ``passes'' homogeneously and distances between simultaneous events can be determined. It is noted that to formalize this intuition, there is no need to invoke affine Euclidean structure (cf. Ref. \cite{Marsden}). 

Given 4-dimensional manifold $\mathcal{V}$, and 3-dimensional manifold $S$, a diffeomorphism
\begin{align}
o: S \times \mathbb{R} \rightarrow \mathcal{V} \nonumber
\end{align}
is called a \emph{slicing}. We define $o_t: S \rightarrow \mathcal{V}; \ x \mapsto o(x,t)$ for each $t \in \mathbb{R}$. The manifold $\mathcal{V}$ is \emph{classical space-time} if
\begin{itemize}
\item[(i)] there is surjective map $\tau: \mathcal{V} \rightarrow \mathbb{R}$, called universal time, with everywhere non-vanishing derivative, such that $S_t := \tau^{-1}(t)$ is diffeomorphic to $S$ for all $t \in \mathbb{R}$
\item[(ii)] there is a slicing $o: S \times \mathbb{R} \rightarrow \mathcal{V}$ which is compatible with time, that is, $o_t(S) = S_t$ for all $t \in \mathbb{R}$
\item[(iii)] there is a symmetric connection on $\mathcal{V}$ by which  $\xdif \tau$ is covariant constant
\item[(iv)] on each $S_t$ there is a Riemannian metric $g_t$, whose Levi-Civita connection $\nabla_t$ coincides with the restriction of the space-time connection to $S_t$.
\end{itemize} 
A slicing $o: S \times \mathbb{R} \rightarrow \mathcal{V}$ of classical space-time is called an \emph{observer}. An observer that is compatible with time (item (ii)) is called \emph{Newtonian observer}. 

Newtonian observer $o: S \times \mathbb{R} \rightarrow \mathcal{V}$ perceives simultaneous events at time $t$ using $o_t:S \rightarrow S_t$. This may be used to carry the metric $g_t$ and connection $\nabla_t$ to $S$. Accordingly, we define the \emph{metric observed by $o$ on $S$ at time $t$} as $g^o_t = o_t^* g_t$, and the \emph{connection observed by $o$ on $S$ at time $t$} as $\nabla_t^o = o_t^* \nabla_t$ (defined using the push-forward by $(o_t^* \nabla_t)_u v = (o_t^{-1})_* \big((\nabla_t)_{(o_t)_* u} (o_t)_*v\big)$). It follows that $\nabla_t^o$ is the Levi-Civita connection of $g^o_t$. 

If there is a fixed (time-independent) metric $g$ on $S$, Newtonian observer $o$ is called \emph{rigid} if $g_t^o = g$ for all $t\in \mathbb{R}$.

To describe material body in space-time, we let $M$ be a reference manifold (with $\mbox{dim}(M) \le 3$), and define \emph{world tube for $M$} as a map
\begin{align}
\Theta: M \times \mathbb{R} \rightarrow \mathcal{V}, \nonumber
\end{align}
such that the map $\Theta_t: M \rightarrow \mathcal{V}; \ X \mapsto \Theta(X,t)$ is an embedding, and fulfills $\Theta_t(M) \subset S_t$, for all $t \in \mathbb{R}$. The path $t \mapsto \Theta(X,t)$ is the \emph{world line for $X$}. The ``absolute'' velocity, or \emph{four-velocity}, of the world line for $X$ is the velocity vector field of the world line, and its ``absolute'' acceleration, or \emph{four-acceleration}, is given by taking the covariant derivative of the four-velocity along the world line, using the space-time connection. Going through all points $X \in M$ gives the four-velocity and four-acceleration of the world tube $\Theta$.

In particular, Newtonian observer $o: S \times \mathbb{R} \rightarrow \mathcal{V}$ is a world tube for $S$, and has a well-defined four-velocity and four-acceleration.

Given a world tube $\Theta: M \times \mathbb{R} \rightarrow \mathcal{V}$ and a Newtonian observer $o: S \times \mathbb{R} \rightarrow \mathcal{V}$, the \emph{motion of $M$ on $S$ relative to $o$} is the map $\theta^o: M \times \mathbb{R} \rightarrow S$ defined by $\Theta(X,t) = o(\theta^o(X,t),t)$. The \emph{(``apparent'') velocity of $\Theta$ relative to $o$} is the velocity vector field of $\theta^o$ (defined by sewing together the velocity vector fields of the paths $t \mapsto \theta^o(X,t)$ corresponding to all $X \in M$. The \emph{(``apparent'') acceleration of $\Theta$ relative to $o$} is given by taking the covariant derivative of the relative velocities of material points along the paths $t \mapsto \theta^o(X,t)$, using the (possibly time dependent) connection observed on $S$ by $o$.

Using this setup, it is straightforward to derive, in particular, the transformation of relative velocities and accelerations under change of observer. These transformations are not given here, for we will restrict the analysis to statics, and allow only \emph{inertial} observers, namely, those whose velocity field is covariant constant and whose acceleration therefore vanishes. The following analysis serves as our opening to the modeling of magneto-elastic coupled problems, and it will need to be extended to allow for dynamics and non-inertial observers. In particular, this would require considering the non-tensorial transformation of accelerations and forces under change of observer (see \cite{Marsden}).

\section{Elastostatics} \label{elastostatics}

In addition to restricting the analysis to small-strain elastostatics, a restriction is made to hyper-elastic constitutive laws, meaning that elastic energy density will be assumed. (It is noted that large amount of work have been made on the geometric analysis of the general elastic problem -- and continuum mechanics in general -- that is free of these restrictive assumptions, see for example \cite{Kanso,Marsden,Segev,Yavari}.) An analysis that restricts to small-strain magneto-hyper-elasticity is justified by the significance of this model in important applications, such as electric motor, sensor, and actuator technologies.

\subsection{Infinitesimal displacement from a reference configuration}

Let us assume that a rigid inertial observer $o: S \times \mathbb{R} \rightarrow \mathcal{V}$ is given, and denote as $g^o$ the observed (time-independent) Riemannian metric on $S$. Further, let us denote as $\theta^o: M \times \mathbb{R} \rightarrow S$ the motion of $M$ in $S$ relative to $o$, and as $\theta^o_0: M \rightarrow S; \ X \mapsto \theta^o(X,0)$ a stress-free reference configuration.

Intuitively, we will be concerned with an infinitesimal displacement of material points from the reference configuration. To formulate the problem on $M$, we consider \emph{displacement} $\nu$ as a vector field on $M$. This is appropriate because we assume both $M$ and $S$ to be 3-dimensional. (More general modelings, where $M$ and $S$ may be of different dimension, can be found in continuum mechanics literature.) This is a slightly atypical choice of a modeling quantity, as the spatial displacement observed by $o$ is now the vector field $(\theta^o_0)_* \nu$ on $\theta^o_0(M)$. In strict formal language, our displacement $\nu$ is a smooth section of the tangent bundle $TM$ of $M$ (with the projection $\pi: TM \rightarrow M$), that is, a smooth map $\nu: M \rightarrow TM$ satisfying $\pi \circ \nu = \mbox{id}_M$ (identity on $M$).

Finally, in preparing to give the formulation on $M$, let us introduce on $M$ the metric $G = (\theta^o_0)^* g^o$ and the connection $\nabla = (\theta^o_0)^* \nabla^o$. It follows that $\nabla$ is the Levi-Civita connection of $G$. Further, the metric $G$ is Euclidean if and only if $g^o$ is Euclidean, because $M$ and $S$ are of the same dimension.

\subsection{Displacement gradient}

In the following we will often work with the category of vector bundles and vector bundle mappings \cite{AMR,AM}. Given two vector bundles $E$ and $E'$ over $M$, we denote as $L(E;E')$ the vector bundle whose fiber above point $X \in M$ is the vector space $L(E_X; E'_X)$ of linear maps from the fiber $E_X$ to the fiber $E'_X$. (For further details, see for example \cite{AMR}.) The (infinite dimensional) vector space of (appropriately smooth) sections of a vector bundle $E$ will be denoted as $\Gamma(E)$.

To express small-strain elasticity and magnetism in analogous form, we bring fore the notion of \emph{displacement gradient}. Intuitively, the displacement gradient serves to describe how the values of displacement are changed to the first order in the neighborhoods of points. To avoid introducing new notation, we will denote the displacement gradient by the symbol $\varepsilon$, which is usually reserved for strain tensor (the symmetric part of displacement gradient, see section \ref{ela_energy_section} below). Formally, the displacement gradient is a section of the vector bundle $L(TM;TM)$, that is, a vector-valued 1-form on $M$. Accordingly, its value at each point $X \in M$ belongs to the vector space $L(T_X M; T_X M)$ of linear maps on the tangent space $T_XM$. 

To formalize the above intuition about the relation of displacement gradient to displacement, we need to compare the displacement vectors at neighboring points of $M$ -- an operation for which we use the connection $\nabla$ on $M$. Since $\nabla$ takes two smooth vector fields $u$ and $v$ to a third smooth vector field $\nabla_u v$, which depends linearly on $v$, and is function-linear\footnote{\label{module1}That is, the mapping $\nabla v: u \mapsto \nabla_u v$ is a structure preserving map on the $\mathcal{F}(M)$-module of vector fields, where $\mathcal{F}(M)$ is the ring of smooth real-valued functions on $M$.} in $u$, we may conceive it as the linear map $\nabla: \Gamma(TM) \rightarrow \Gamma(L(TM;TM)); \ v \mapsto \nabla v$. The relation of displacement gradient $\varepsilon$ to displacement $\nu$ is given as
\begin{equation} \label{strain}
\varepsilon = \nabla \nu.
\end{equation}

The above relation is invariant to changes of observer, provided that the observed connection is used. By this we mean the following. The change of observer corresponds to a diffeomorphism $\chi: M \rightarrow M$. The metric and connection on $M$ that correspond to the new observer are $\hat{G} = (\chi^{-1})^* G$ and $\hat{\nabla} = (\chi^{-1})^* \nabla$. Denoting as $\hat{\varepsilon}$ and $\hat{\nu}$ the transformed displacement gradient and displacement, that is, $\hat{\varepsilon} = \chi_* \circ \varepsilon \circ \chi_* ^{-1}$ and $\hat{\nu} = \chi_* \nu$, we have $\hat{\varepsilon} = \hat{\nabla} \hat{\nu}$. 

\subsection{Force functional} \label{force}

We recall that when $E$ and $E'$ are vector bundles, with projections $\pi:E \rightarrow B$ and $\pi':E' \rightarrow B'$, a vector bundle map $h: E \rightarrow E'$ induces a unique map $h_B: B \rightarrow B'$ such that $\pi' \circ h = h_B \circ \pi$ holds \cite{AMR}. It is said that $h$ is a vector bundle map over $h_B$.

In the following, we will make use of a correspondence between smooth sections of $L(E;E')$ and vector bundle maps $E \rightarrow E'$ over the identity on $M$. This correspondence is bijective, associating to a section $s$ of $L(E;E')$ the vector bundle map $\hat{s}: E \rightarrow E'$ over the identity on $M$ whose restriction to the fiber $E_X$ is defined by $\hat{s}_X = s(X)$ for all $X \in M$, and vice versa. In the following, when we say that a section $s$ of $L(E;E')$ operates on a section $s'$ of $E$ we mean the map $s' \mapsto \hat{s} \circ s'$, taking $s'$ to a section of $E'$.

Forces inside materials operate on virtual displacement vector fields to produce 3-forms that can be integrated to yield contributions to virtual work. Accordingly, we introduce \emph{body force} as a section of $L(TM; \tilde{\bigwedge}^3(T M))$, where $\tilde{\bigwedge}^3(T M)$ is the bundle of 3-covectors on $M$. We use ``tilde'' in the notation as a reminder that one should really deal with twisted forms \cite{Frankel,Kovanen,Burke}. (Twisted forms are treated with varying accuracy in this paper. If $M$ is orientable, and a specific orientation is selected, this issue may be omitted, as twisted p-forms become represented by p-forms.) With hindsight, we will denote the body force as $\tilde{f} \dot{\wedge}$. For a virtual displacement vector field $\delta \nu$ the body force $\tilde{f} \dot{\wedge}$ produces the 3-form $\tilde{f} \dot{\wedge} \delta \nu$, which may be integrated over $M$ (independently of orientation, or even when $M$ is not orientable, using integration of twisted forms).

In addition to body forces we allow surface forces on the boundary $\partial M$ of $M$. Using the natural inclusion $i: \partial M \rightarrow M$ we may consider the pull-back bundle $i^* (TM)$. This is the vector bundle whose base manifold is $\partial M$ and fiber above $X \in \partial M$ is (identified with) the tangent space $T_{i(X)} M$. We take \emph{surface force} $\tilde{\tau} \dot{\wedge}$ as a section of $L(i^*(TM); \tilde{\bigwedge}^2(T \partial M))$, where $\tilde{\bigwedge}^2(T \partial M)$ is the bundle of 2-covectors on $\partial M$. Accordingly, surface force can operate on sections of $i^*(TM)$ to produce 2-forms on $\partial M$. Finally, for a section $\delta \nu$ of $TM$, we denote as $i^*\delta \nu$ the section of $i^*(TM)$ defined by $i^* \delta \nu = \delta \nu \circ i$.

Global force should give the virtual work done under virtual displacements of the body $M$. Accordingly, given body force $\tilde{f} \dot{\wedge}$ and surface force $\tilde{\tau} \dot{\wedge}$, we consider \emph{force} $F$ as the linear functional on $\Gamma(TM)$ defined using integration as
\begin{equation} \label{global_force_def}
F(\delta \nu) = \int_M \tilde{f} \dot{\wedge} \delta \nu + \int_{\partial M} \tilde{\tau} \dot{\wedge} i^* \delta \nu
\end{equation}
for all $\delta \nu \in \Gamma(TM)$. The surface force may not be given beforehand on some part of $\partial M$, and then it will be defined on this part using the notion of stress (section \ref{equilibrium_eq}). Stress will be defined in this paper through a constitutive law (section \ref{stress}), although it could be introduced without taking sides to the constitutive law, see for example \cite{Segev}.

The relation \eqref{global_force_def} is invariant to changes of \emph{inertial} observers. (For the general transformation of forces, see \cite{Marsden}.)

\subsection{Elastic energy and invariance to rigid displacements} \label{ela_energy_section}

In the context of small-strain and hyper-elasticity assumptions, an elastic energy density is assumed, which depends on the displacement only through the symmetric part of the displacement gradient. This means that rigid infinitesimal displacements will not affect elastic energy. When the space manifold is Euclidean, such displacements are translations, rotations, and their combinations.

Let $E$ and $E'$ be vector bundles over $M$ and $h: E \rightarrow E'$ a vector bundle map, that is, a smooth fiber preserving map which is linear in each fiber. This allows us to define a linear map $\Gamma(h): \Gamma(E) \rightarrow \Gamma(E'); \ s \mapsto h \circ s$. The symbol $\Gamma$ represents a functor from the category of vector bundles over $M$ to the category of vector spaces. 

In the following, $h$ will be a vector bundle map over the identity on $M$. It will be constructed by using linear maps $h_X: E_X \rightarrow E_X'$ defined identically for all $X \in  M$ (such that the resulting map $h$ will be smooth). Denoting as $\pi: E \rightarrow M$ and $\pi': E' \rightarrow M$ the bundle projections, the map $h$ is defined by $h(e) = h_X(e) \in (\pi')^{-1}(X)$ for all $e \in \pi^{-1}(X)$ and all $X \in M$. Often the maps $h_X$ will be bijections, making $h$ a vector-bundle isomorphism \cite{AMR}, and $\Gamma(h)$ a linear isomorphism.

Let us introduce \emph{elastic energy density} as a fiber bundle morphism $\tilde{\Psi}: L(TM;TM) \rightarrow \tilde{\bigwedge}^3(T M)$ over the identity on $M$. This means $\tilde{\Psi}$ is a smooth fiber preserving map, that is, it maps vectors from $L(T_XM;T_XM)$ to $\tilde{\bigwedge}^3(T_{\mbox{id}_M(X)} M)$ for all $X \in M$. (See for example \cite{AMR}.) Therefore, its restriction to the fiber $L(T_XM;T_XM)$ is a smooth map $\tilde{\Psi}_X: L(T_XM; T_XM) \rightarrow \tilde{\bigwedge}^3(T_X M)$ for each $X \in M$. In the most simple cases (corresponding to linear constitutive laws) these restricted maps will be quadratic (as in section \ref{example} below). Taking the composition of $\tilde{\Psi}$ and a section $\varepsilon$ of $L(TM;TM)$ we get a section of $\tilde{\bigwedge}^3(T M)$, and then we will say that $\tilde{\Psi}$ operates on $\varepsilon$ to produce the 3-form $\tilde{\Psi}(\varepsilon)$.

To express the invariance to rigid displacements, let us decompose the displacement gradient into symmetric and antisymmetric parts using the metric $G$. For this, we define linear maps $\mbox{sym}: L(T_XM;T_XM) \rightarrow L(T_XM;T_XM)$ identically for all $X \in M$, by setting $\mbox{sym}(\varepsilon)^i_j = (\varepsilon^i_j + G_{lj}G^{ik} \varepsilon^l_k)/2$. For the antisymmetric (or skew) part, we define linear maps $\mbox{skw}: L(T_XM;T_XM) \rightarrow L(T_XM;T_XM)$ by $\mbox{skw}(\varepsilon)^i_j = (\varepsilon^i_j - G_{lj}G^{ik} \varepsilon^l_k)/2$. Then, for any $\varepsilon \in L(T_XM;T_XM)$ we have the decomposition $\varepsilon = \mbox{sym}(\varepsilon) + \mbox{skw}(\varepsilon)$. The resulting vector bundle maps, and the linear maps given by $\Gamma$, will be denoted by the same symbols.

The classical small-strain tensor, defined on $\theta^o_0(M) \subset S$ using the rigid observer's metric, is $\frac{1}{2}\mathcal{L}_{(\theta^o_0)_* \nu} g^o$, where $\mathcal{L}$ is the Lie derivative. Its relation to the displacement gradient is given by using $\frac{1}{2}(\mathcal{L}_\nu G)_{ij} = G_{ik}\mbox{sym}(\nabla \nu)^k_j$ and the naturality of the Lie derivative with respect to push-forward, that is, $\mathcal{L}_\nu G = (\theta^o_0)^* \big(\mathcal{L}_{(\theta^o_0)_* \nu} g^o\big)$. In the present formulation, the invariance of elastic energy to rigid displacements is taken into account by insisting $\tilde{\Psi}(\nabla \nu)$ not to depend on $\mbox{skw}(\nabla \nu)$.

Elastic energy density $\tilde{\Psi}$ lets us introduce \emph{elastic energy} as the map 
\begin{align}
W: \Gamma(TM) \rightarrow \mathbb{R}; \ \nu \mapsto \int_M \tilde{\Psi}(\nabla \nu), \label{ela_energy}
\end{align}
where $\tilde{\Psi}$ is considered as a (non-linear) map $\Gamma(L(TM;TM)) \rightarrow \Gamma(\tilde{\bigwedge}^3(T M))$, and $\int_M$ is a linear map $\Gamma(\tilde{\bigwedge}^3(T M)) \rightarrow \mathbb{R}$. This expression for the elastic energy is invariant to changes of observer, provided that the observed connection is used. The transformed elastic energy density under $\chi: M \rightarrow M$ is defined, using the pull-back of 3-forms, such that $\hat{\tilde{\Psi}}(\hat{\varepsilon}) = (\chi^{-1})^*\big(\tilde{\Psi}(\chi_*^{-1} \circ \hat{\varepsilon} \circ \chi_*\big)$, and the elastic energy is given by integrating $\tilde{\Psi}(\nabla \nu) = \chi^* \big(\hat{\tilde{\Psi}}(\hat{\nabla} \hat{\nu})\big)$ over $M$. Therefore, the variational procedure of the next section may be carried out in case of arbitrary observer that uses the observed connection.\footnote{For an \emph{arbitrary connection} to be allowed, we should allow the elastic energy density to depend also on the point values of displacement---not only on the derivatives of displacement. For example, it may be useful to adjust the connection to a particular body configuration. Let $\nabla^{\mbox{ad}}$ denote an arbitrary ``adjusted'' connection of the observer whose observed connection on $M$ is $\nabla$. Because the difference $\nabla - \nabla^{\mbox{ad}}$ transforms as a tensor, it is possible to define an ``adjusted'' elastic energy density $\tilde{\Psi}^{\mbox{ad}}: L(TM;TM) \oplus TM \rightarrow \tilde{\bigwedge}^3(T M)$, such that $\tilde{\Psi}^{\mbox{ad}}(\nabla^{\mbox{ad}} \nu, \nu) = \tilde{\Psi}(\nabla \nu)$. This suggests that there is a connection-free manner of representing elastic energy and its variations, using first order jets of sections of $TM$. We will not pursue this point further in this paper.}

\subsection{Variational formulation}

Elastic energy variations appear through variations of elastic energy density. For the variations of elastic energy density that result from variations of displacement gradient, we take the derivatives of the restricted maps $\tilde{\Psi}_X: L(T_XM; T_XM) \rightarrow \tilde{\bigwedge}^3(T_X M)$ corresponding to all $X \in M$. At each point $X \in M$, the derivative 
\begin{align}
\xDif(\tilde{\Psi}_X): L(T_XM;T_XM) \rightarrow L(L(T_XM;T_XM); \tilde{\bigwedge}^3(T_X M)) \nonumber
\end{align}
has the operation 
\begin{align}
\xDif(\tilde{\Psi}_X): \varepsilon(X) \mapsto \xDif(\tilde{\Psi}_X)(\varepsilon(X)). \nonumber
\end{align}
The derivative depends only on the topologies on the spaces $L(T_XM;T_XM)$ and $\tilde{\bigwedge}^3(T_X M)$; not on any particular norms that induce the topologies \cite{AMR}. Further, arbitrary norms may be used here, as in case of a finite dimensional vector space (over real or complex numbers) all norms induce the same topology \cite{AMR}. We may define, for each displacement gradient value $\varepsilon$ (section of $L(TM;TM)$), a section $D \tilde{\Psi}(\varepsilon)$ of the bundle $L(L(TM;TM); \tilde{\bigwedge}^3(T M))$ by
\begin{align} \label{deriv_ela_energy_dens}
D \tilde{\Psi}(\varepsilon)(X) = \xDif(\tilde{\Psi}_X)(\varepsilon(X)),
\end{align}
for all $X \in M$. The section $D \tilde{\Psi}(\varepsilon)$ can operate on variations of displacement gradient (sections of $L(TM;TM)$) to yield variations of elastic energy density (sections of $\tilde{\bigwedge}^3(T M)$).

Now we are in the position to express a variational principle as the defining equation for the displacement $\nu$, when the energy density $\tilde{\Psi}$, the body force $\tilde{f} \dot{\wedge}$, and surface force $\tilde{\tau} \dot{\wedge}$ on some (possibly empty) part of $\partial M$, are specified. If $\tilde{\tau} \dot{\wedge}$ is not specified at some part $S_e \subset \partial M$ we specify $i^*\nu$ on this boundary part. Then, according to the basic principle, we require that the virtual deformation work done by the given forces coincides with the variation of elastic energy for all kinematically admissible virtual displacements. Accordingly\footnote{\label{Elastic_energy_note}Variations of elastic energy may be considered once a norm is given for the vector space $\Gamma(TM)$, and its completion to a Banach space is performed. A norm may be defined, for instance, by using the metric $G$ on $M$ to define an inner product for $\Gamma(TM)$. Then, the variation of elastic energy corresponding to virtual displacement $\delta \nu \in \Gamma(TM)$ is given by using the Fr\'echet derivative as $\xDif W(\nu) \cdot \delta \nu$ (assuming the differentiability of $W$). Let us assume that norms are given also for $\Gamma(L(TM;TM))$ and $\Gamma(\tilde{\bigwedge}^3(T M))$ by similar procedure, and that their completion is performed to obtain Banach spaces. (Inner product for $\Gamma(\tilde{\bigwedge}^3(T M))$ may be defined by $\langle \cdot, \cdot \rangle = \int_M \cdot \wedge \star \cdot$, using the exterior product $\wedge$ of differential forms and the Hodge operator $\star$ implied by the metric $G$. Similarly, inner products for both $\Gamma(TM)$ and $\Gamma(L(TM;TM))$ may be defined by $\langle \cdot, \cdot \rangle = \int_M \cdot \dot{\wedge} \star^{\flat} \cdot$, using the product $\dot{\wedge}$ and the Hodge operator $\star^{\flat}$ defined later in this section.) Then, by using the chain rule, and taking into account the linearity of $\nabla$ and $\int_M$, the variation of elastic energy becomes $\int_M \xDif \tilde{\Psi}(\nabla \nu) \nabla \delta \nu$ (assuming the required differentiability). It is not difficult to show, assuming differentiability, that the Fr\'echet derivative of $\tilde{\Psi}: \Gamma(L(TM;TM)) \rightarrow \Gamma(\tilde{\bigwedge}^3(T M))$ coincides with the (pointwise) derivative defined in \eqref{deriv_ela_energy_dens}. Finally, for \emph{arbitrary metric} to be allowed in this variational procedure, it is left to be shown that the differentiability of $\nabla$, $\tilde{\Psi}$, and $\int_M$, is independent of the metric conferred on $M$ by which the required norms are constructed.}, the problem is to find $\nu \in \Gamma(TM)$, with $i^* \nu$ predefined on $S_e \subset \partial M$, such that
\begin{equation} \label{var_princip}
\int_M D \tilde{\Psi}(\nabla \nu)\nabla \delta \nu = F(\delta \nu)
\end{equation}
for all $\delta \nu \in \Gamma(TM)$, with $i^* \delta \nu$ zero on $S_e$. On the right-hand-side the virtual deformation work $F(\delta \nu)$ is evaluated according to \eqref{global_force_def}. 

In equation \eqref{var_princip}, the following requirement concerns $F$, when $\tilde{\tau}$ is predefined on all of $\partial M$. Because of the invariance of elastic energy density to rigid displacements, this prescribed load must yield zero virtual work for rigid virtual displacements $\delta \nu$, that is, those satisfying $\mbox{sym}(\nabla \delta \nu) = 0$. Indeed, the energy density $\tilde{\Psi}$ is of the form $\tilde{\Psi}' \circ \mbox{sym}$ for some fiber bundle morphism $\tilde{\Psi}': L(TM;TM) \rightarrow \tilde{\bigwedge}^3(T M)$, and hence, using the chain rule to the pointwise derivative, we have
\begin{align}
D \tilde{\Psi}(\nabla \nu) \nabla \delta \nu = D \tilde{\Psi}'(\mbox{sym}(\nabla \nu) \mbox{sym}(\nabla \delta \nu), \nonumber
\end{align}
such that rigid virtual displacements produce zero elastic energy variations.

\subsection{Local forces as covector-valued forms} \label{local_forces}

Given a vector bundle $E$ over $M$, its dual bundle $L(E, M \times \mathbb{R})$ will be denoted as $E^*$. The specific notation $T^*M$ will be used for the cotangent bundle, that is, the dual bundle of $TM$. We will sometimes abuse the notation by using the same symbol for a section of a bundle and its value at a point.

Body force may be viewed as a covector-valued 3-form because of an isomorphism between $L(TM; \tilde{\bigwedge}^3(T M))$ and $L(\tilde{\bigwedge}_3(T M); T^*M)$, where $\tilde{\bigwedge}_3(T M) = \big(\tilde{\bigwedge}^3(T M)\big)^*$ is the bundle of 3-vectors on $M$. We define the linear isomorphism between $L(T_XM; \tilde{\bigwedge}^3(T_X M))$ and $L(\tilde{\bigwedge}_3(T_X M); T_X^*M)$ identically for all $X \in M$. The isomorphism assigns to the element $\tilde{f}$ of $L(\tilde{\bigwedge}_3(T_X M); T_X^*M)$ the element $\tilde{f} \dot{\wedge} \in L(T_XM; \tilde{\bigwedge}^3(T_X M))$ defined by
\begin{equation}
(\tilde{f} \dot{\wedge} v) (\tilde{u}) = \tilde{f}(\tilde{u})(v)
\end{equation}
for all $v \in T_XM$ and $\tilde{u}  \in \tilde{\bigwedge}_3(T_X M)$. Accordingly, body force may also be considered as the section $\tilde{f}$ of $L(\tilde{\bigwedge}_3(T M); T^*M)$. The generalization of this isomorphism to covector-valued p-forms is straightforward \cite{Segev,Kovanen}. Similarly, surface force may be considered as the section $\tilde{\tau}$ of $L(\tilde{\bigwedge}_2(T \partial M); i^*(T^*M))$, where $\tilde{\bigwedge}_2(T \partial M) = \big(\tilde{\bigwedge}^2(T \partial M)\big)^*$ is the bundle of 2-vectors on $\partial M$, and $i^*(T^*M)$ is the pull-back of $T^*M$ by $i$.

\subsection{Stress} \label{stress}

Here, we will define stress as a covector-valued 2-form on $M$, that is, as a section of $L(\tilde{\bigwedge}_2(T M); T^*M)$, which will be able to produce energy density variations from given displacement gradient variations. This performance will be obtained by using an isomorphism between $L(\tilde{\bigwedge}_2(T M); T^*M)$ and $L(L(TM;TM); \tilde{\bigwedge}^3(T M))$. This will allow us to consider stress as a section of $L(L(TM;TM); \tilde{\bigwedge}^3(T M))$, which can operate on sections of $L(TM; TM)$ to produce sections of $\tilde{\bigwedge}^3(T M)$. 

We define the linear isomorphism between $L(\tilde{\bigwedge}_2(T_X M); T_X^*M)$ and $L(L(T_XM;T_XM); \tilde{\bigwedge}^3(T_X M))$ identically for all $X \in M$. The isomorphism assigns to $\tilde{\sigma} \in L(\tilde{\bigwedge}_2(T_X M); T_X^*M)$ the element $\tilde{\sigma} \dot{\wedge} \in L(L(T_XM;T_XM); \tilde{\bigwedge}^3(T_X M))$ defined by
\begin{equation} \label{ext_product}
(\tilde{\sigma} \dot{\wedge} e) (u,v,w) = \tilde{\sigma}(u,v)(e(w)) + \tilde{\sigma}(w,u)(e(v)) + \tilde{\sigma}(v,w)(e(u))
\end{equation}
for all $e \in L(T_XM;T_XM)$ and $u,v,w \in T_XM$. Linearity and bijectivity of this map is verified in Appendix \ref{isomorphism}. Since the above deals with twisted forms the triplet $(u,v,w)$ on the left hand side is in fact a twisted 3-vector, and it therefore comes with a transverse orientation. In our three dimensional case this is just a sign (plus or minus). The tuples $(u,v)$, $(w,u)$ and $(v,w)$ on the right hand side are then twisted 2-vectors, whose transverse orientations (crossing directions) are specified by $\pm w$, $\pm v$ and $\pm u$, respectively, the sign being that of $(u,v,w)$. 

The given isomorphism between the bundles $L(\tilde{\bigwedge}_2(T M); T^*M)$ and $L(L(TM;TM); \tilde{\bigwedge}^3(T M))$ is a special case of the relation between ``variational stresses'' and ``Cauchy stresses'' given in \cite{Segev2}.

We may now define \emph{stress} $\tilde{\sigma}$ as the section of $L(\tilde{\bigwedge}_2(T M); T^*M)$ whose corresponding section $\tilde{\sigma} \dot{\wedge}$ of the bundle $L(L(TM;TM); \tilde{\bigwedge}^3(T M))$ coincides with the derivative of elastic energy density, that is
\begin{equation} \label{const}
\tilde{\sigma} \dot{\wedge} = D \tilde{\Psi}(\varepsilon),
\end{equation}
which serves as a \emph{constitutive law} between stress and displacement gradient. (Again, we note that it is possible to define stress in the general context, without taking sides to the constitutive law, see for example \cite{Segev}.)

The constitutive law \eqref{const} is invariant to changes of observer, that is, we have $\hat{\tilde{\sigma}} \dot{\wedge} = D \hat{\tilde{\Psi}}(\hat{\varepsilon})$, where $\hat{\tilde{\sigma}} = (\chi^{-1})^* \circ \tilde{\sigma} \circ \chi^{-1}_*$. (Here, we use the push-forward of bivectors, and pull-back of 1-forms.) This can be verified by considering the transformation rule of $\tilde{\Psi}$, applying the chain rule to the pointwise derivative, and making use of the naturality of the above defined isomorphism with respect to diffeomorphisms, that is, $\tilde{\sigma} \dot{\wedge} \delta \varepsilon = \chi^* (\hat{\tilde{\sigma}} \dot{\wedge} \delta \hat{\varepsilon})$.

\subsection{Natural product for vector- and covector-valued forms} \label{product}

Let us now define a product for vector- and covector-valued forms which has similar properties as the exterior product of differential forms. This will motivate the above usage of the symbol $\dot{\wedge}$. The product appears also in \cite{Kanso,Yavari,Frankel,Kovanen}.

We first define bilinear maps
\begin{align}
\dot{\wedge}:L(\bigwedge_p(T_X M); T_X^*M) \times L(\bigwedge_q(T_X M); T_XM) \rightarrow \bigwedge^{p+q}(T_X M), \nonumber
\end{align}
identically for all $X \in M$. Let us denote as $P(p,q)$ the set of all permutations $\sigma$ of the index set $\{1,\dots,p+q\}$ that satisfies $\sigma(1) < \cdots < \sigma(p)$ and  $\sigma(p+1) < \cdots < \sigma(q)$. Then, for $\omega \in L(\bigwedge_p(T_X M); T_X^*M)$ and $\eta \in L(\bigwedge_q(T_X M); T_XM)$, their product $\omega \dot{\wedge} \eta \in \bigwedge^{p+q}(T_X M)$ is defined, for $p+q \le \mbox{dim}(M)$, by 
\begin{align} \label{product1}
\omega \dot{\wedge} \eta (v_1,\dots,v_{p+q}) = \sum_{\sigma \in P(p,q))} \mbox{sgn}(\sigma)\omega(v_{\sigma(1)},\dots,v_{\sigma(p)})(\eta(v_{\sigma(p+1)},\dots,v_{\sigma(p+q)})),
\end{align}
for all $v_1,\dots,v_{p+q} \in T_XM$, where $\mbox{sgn}(\sigma)$ is the signature of the permutation. This defines the values of $\omega \dot{\wedge} \eta$ for simple (p+q)-vectors. This is sufficient when $\mbox{dim}(M) \le 3$.

The bilinear map
\begin{align}
\dot{\wedge}: L(\bigwedge_q(T_X M); T_XM) \times L(\bigwedge_p(T_X M); T_X^*M) \rightarrow \bigwedge^{p+q}(T_X M) \nonumber
\end{align}
is defined similarly to fullfill graded anticommutativity
\begin{align}
\omega \dot{\wedge} \eta = (-1)^{pq} \eta \dot{\wedge} \omega,
\end{align}
in analogy to the exterior product of differential forms.

By the above definition, we have a vector bundle map over the identity on $M$ from the bundle $L(\bigwedge_p(T M); T^*M)$ to $L(L(\bigwedge_q(T M); TM); \bigwedge^{p+q}(T M))$, with the operation $\omega \mapsto \omega \dot{\wedge}$. Also, we have a vector bundle map from $L(\bigwedge_q(T M); TM)$ to $L(L(\bigwedge_p(T M); T^*M); \bigwedge^{p+q}(T M))$ over the identity on $M$. By using these maps we may construct the products
\begin{align}
\dot{\wedge}: \Gamma(L(\bigwedge_p(T M); T^*M)) \times \Gamma(L(\bigwedge_q(T M); TM)) \rightarrow \Gamma(\bigwedge^{p+q}(T M)) \nonumber
\end{align}
and 
\begin{align}
\dot{\wedge}: \Gamma(L(\bigwedge_q(T M); TM)) \times \Gamma(L(\bigwedge_p(T M); T^*M)) \rightarrow \Gamma(\bigwedge^{p+q}(T M)) \nonumber
\end{align}
that are bilinear and fullfill graded anticommutativity. Similar construction applies in case of forms defined on $\partial M$ (and is required to deal with traces of forms).

\subsection{Traces} \label{traces}

We will sometimes regard sections of $TM$ as vector-valued 0-forms, that is, as sections of the bundle $L(\bigwedge_0(T M);TM)$, where $\bigwedge_0(T M)$ is defined to be $M \times \mathbb{R}$. This is simply achieved by assigning to $u \in \Gamma(TM)$ the element of $\Gamma(L(\bigwedge_0(T M);TM))$ whose operation on $1 \in \Gamma(M \times \mathbb{R})$ produces $u$.

The restriction of forms to the boundary $\partial M$ may be defined by using the natural inclusion $i: \partial M \rightarrow M$. For a vector-valued 0-form $v \in \Gamma(L(\bigwedge_0(T M);TM))$ its trace $\mbox{t} v$ is the element of $\Gamma(L(\bigwedge_0(T \partial M); i^*(TM)))$ defined by $\mbox{t} v (1) = v(1) \circ i$. The trace defined here replaces, when we identify vector fields with vector-valued 0-forms, the restriction of vector fields defined in \ref{force}. (The required commutative diagram is readily checked.)

To define traces of covector-valued p-forms, we make use of the following procedure. For the element $w \in \Gamma(L(\bigwedge_0(T \partial M); i^*(TM)))$ we denote as $l(w)$ an element of $\Gamma(L(\bigwedge_0(T M);TM))$ such that $\mbox{t} (l(w)) = w$. Then, the trace of covector-valued p-form $\omega \in \Gamma(L(\bigwedge_p(T M); T^*M))$ is the element $\mbox{t} \omega \in \Gamma(L(\bigwedge_p(T \partial M); i^*(T^*M)))$ defined, for $p < \mbox{dim}(M)$, by
\begin{align} \label{trace}
\mbox{t} \omega \dot{\wedge} w = \mbox{t}(\omega \dot{\wedge} l(w))
\end{align}
for all $w \in \Gamma(L(\bigwedge_0(T \partial M); i^*(TM)))$, where $\mbox{t}$ on the right-hand-side is the trace of real-valued p-forms defined as the pull-back by $i$. (To pull back twisted forms, $\partial M$ must be transverse orientable \cite{Frankel}.) From \eqref{trace}, it follows that $\mbox{t}(\omega \dot{\wedge} v) = \mbox{t}\omega \dot{\wedge} \mbox{t} v$. The trace of vector-valued p-forms may be defined similarly (but is not required in this paper).

\subsection{Equilibrium equation} \label{equilibrium_eq}

Assuming the stress form $\tilde{\sigma}$ smooth enough, the variational (or weak) formulation \eqref{var_princip} has the equilibrium equation with boundary conditions as a pointwise (or strong) counterpart. We go through the derivation here, as it will be a feasible introduction to the covariant exterior derivative. We will sometimes view vector fields as vector-valued 0-forms, and vice versa, without making the difference explicit in the notation. 

Considering \eqref{var_princip}, with the definitions \eqref{global_force_def} and \eqref{const}, and with the term $\int_{\partial M} \mbox{t} (\tilde{\sigma} \dot{\wedge} \delta\nu)$ added and subtracted. We have
\begin{equation}
\int_M \tilde{f} \dot{\wedge} \delta \nu + \int_{\partial M} \mbox{t} (\tilde{\sigma} \dot{\wedge} \delta\nu) - \int_M \tilde{\sigma} \dot{\wedge} \nabla \delta \nu + \int_{\partial M} \tilde{\tau} \dot{\wedge} \mbox{t} \delta\nu - \int_{\partial M} \mbox{t} (\tilde{\sigma} \dot{\wedge} \delta\nu) = 0
\end{equation}
for all $\delta \nu \in \Gamma(L(\bigwedge_0(T M);TM))$, with $\mbox{t} \delta \nu$ zero on $S_e \subset \partial M$. Let us express the integrand in the rightmost term as $\mbox{t} \tilde{\sigma} \dot{\wedge} \mbox{t} \delta\nu$, and replace the other boundary integral involving $\mbox{t} (\tilde{\sigma} \dot{\wedge} \delta\nu)$ by an integral over $M$ by using the exterior derivative $\xdif$. We have
\begin{equation} \label{weak_ela}
\int_M \tilde{f} \dot{\wedge} \delta \nu + \int_M (\xdif(\tilde{\sigma} \dot{\wedge} \delta\nu) - \tilde{\sigma} \dot{\wedge} \nabla \delta \nu) + \int_{\partial M} (\tilde{\tau} - \mbox{t} \tilde{\sigma}) \dot{\wedge} \mbox{t} \delta\nu = 0
\end{equation}
for all $\delta \nu \in \Gamma(L(\bigwedge_0(T M);TM))$, with $\mbox{t} \delta \nu$ zero on $S_e$. The term $\xdif(\tilde{\sigma} \dot{\wedge} \delta\nu)$ contains derivatives of both $\tilde{\sigma}$ and $\delta \nu$, but it turns out that the involved derivatives of $\delta \nu$ are canceled when we subtract the term $\tilde{\sigma} \dot{\wedge} \nabla \delta \nu$. Therefore, the map $\delta \nu \mapsto \xdif(\tilde{\sigma} \dot{\wedge} \delta\nu) - \tilde{\sigma} \dot{\wedge} \nabla \delta \nu$ is not only linear but function-linear\footnote{That is, it is a structure preserving map on the $\mathcal{F}(M)$-module of vector fields, where $\mathcal{F}(M)$ is as in footnote \ref{module1}.}, and we have the remarkable situation that this map defines a section of $L(TM; \tilde{\bigwedge}^3(T M))$. Further, since $\Gamma(L(TM; \tilde{\bigwedge}^3(T M)))$ is isomorphic to $\Gamma(L(\tilde{\bigwedge}_3(T M), T^*M))$, there is a unique covector-valued 3-form, denoted here as $\xdif_{\nabla} \tilde{\sigma}$, such that  
\begin{equation} \label{cov_ext_der}
\xdif_{\nabla} \tilde{\sigma} \dot{\wedge} \delta \nu = \xdif(\tilde{\sigma} \dot{\wedge} \delta\nu) - \tilde{\sigma} \dot{\wedge} \nabla \delta \nu
\end{equation}
for all $\delta \nu \in \Gamma(TM)$. Using this in (\ref{weak_ela}), we get
\begin{equation}
\int_M (\tilde{f} + \xdif_{\nabla} \tilde{\sigma}) \dot{\wedge} \delta \nu + \int_{\partial M} (\tilde{\tau} - \mbox{t} \tilde{\sigma}) \dot{\wedge} \mbox{t} \delta\nu = 0
\end{equation}
for all $\delta \nu \in \Gamma(L(\bigwedge_0(T M);TM))$, with $\mbox{t} \delta \nu$ zero on $S_e$. Thus, with variations vanishing on $\partial M$, we get\footnote{\label{innerproduct_property}Here, we may use a metric on $M$ to define (overloading the notation $M$ for the manifold $M \setminus \partial M$) an inner product on $\Gamma(L(\tilde{\bigwedge}_3(T M), T^*M))$ by $\langle \cdot, \cdot \rangle = \int_M \cdot \dot{\wedge} \star^{\sharp} \cdot$, employing the Hodge operator $\star^{\sharp}$ of subsection \ref{Hodge}. We then have the fact that $\langle \tilde{f} + \xdif_{\nabla} \tilde{\sigma}, \star^{\flat} \delta \nu \rangle = 0$ for all $\star^{\flat} \delta \nu \in \Gamma(L(\tilde{\bigwedge}_3(T M), T^*M))$ implies $\tilde{f} + \xdif_{\nabla} \tilde{\sigma} = 0$.} the pointwise equilibrium equation
\begin{equation} \label{point_equi}
-\xdif_{\nabla} \tilde{\sigma} = \tilde{f},
\end{equation}
and with variations supported on $\partial M \setminus S_e$, we get\footnote{Using appropriate inner product, as in footnote \ref{innerproduct_property}.} the boundary condition
\begin{equation} 
\mbox{t} \tilde{\sigma} = \tilde{\tau} \ \ \mbox{on} \ \ \partial M \setminus S_e.
\end{equation}

Finally, to define the force functional of section \ref{force}, we set
\begin{align}
\tilde{\tau} = \mbox{t} \tilde{\sigma}  \ \ \mbox{on} \ \ S_e.
\end{align}

\subsection{Covariant exterior derivative}

The above defines operator $\xdif_{\nabla}: \Gamma(L(\tilde{\bigwedge}_2(T M); T^*M)) \rightarrow \Gamma(L(\tilde{\bigwedge}_3(T M); T^*M))$, which is linear by the linearity of exterior derivative and bilinearity of the product $\dot{\wedge}$. This operator may be generalized to vector- and covector-valued p-forms as follows. The operator defined here is a special case of the derivatives defined in \cite{Segev2,Segev3}. See also \cite{Kanso,Yavari,Frankel,Kovanen}. For covector-valued p-forms the covariant exterior derivative is the linear map 
\begin{align}
\xdif_{\nabla}: \Gamma(L(\bigwedge_p(TM); T^*M)) \rightarrow \Gamma(L(\bigwedge_{p+1}(T M); T^*M))
\end{align}
defined by
\begin{align}
\xdif_{\nabla} \omega \dot{\wedge} u = \xdif(\omega \dot{\wedge} u) - (-1)^p \omega \dot{\wedge} \nabla u
\end{align}
for all $\omega \in \Gamma(L(\bigwedge_p(T M); T^*M))$ and $u \in \Gamma(TM)$. In symmetric fashion, for vector-valued p-forms the covariant exterior derivative is the linear map
\begin{align}
\xdif_{\nabla}: \Gamma(L(\bigwedge_p(TM); TM)) \rightarrow \Gamma(L(\bigwedge_{p+1}(TM); TM))
\end{align}
defined by
\begin{align} \label{cov_ext_der_vec}
\xdif_{\nabla} \eta \dot{\wedge} \alpha = \xdif(\eta \dot{\wedge} \alpha) -  (-1)^p \eta \dot{\wedge} \xdif_{\nabla} \alpha
\end{align}
for all $\eta \in \Gamma(L(\bigwedge_p(TM); TM))$ and $\alpha \in \Gamma(L(\bigwedge_0(T M); T^*M))$. 

In case of a vector-valued 0-form the derivative of (\ref{cov_ext_der_vec}) reduces to the covariant derivative of the corresponding vector field. This is a straightforward consequence of the above definitions. Therefore, as a replacement for \eqref{strain} we may write
\begin{equation} \label{strain_2}
\varepsilon = \xdif_{\nabla} \nu
\end{equation}
where $\nu \in \Gamma(L(T^0M; TM))$.

The derivative $\xdif_{\nabla}$ can be used to express curvature, as for vector fields $u,v,w$ we have
\begin{align}
(\xdif_{\nabla} \xdif_{\nabla} u)(v,w) = R(v,w)u, 
\end{align}
where $R$ is the curvature of the connection \cite{Frankel}. Thus, in case of a flat connection we have $\xdif_{\nabla} \xdif_{\nabla} u = 0$ for all $u \in \Gamma(L(\bigwedge_0(T M); TM))$. In this flat case we further have $\xdif_{\nabla} \xdif_{\nabla} = 0$ for general vector- and covector-valued p-forms, in analogy to the property $\xdif \xdif = 0$ of the exterior derivative $\xdif$.

\subsection{Hodge operators} \label{Hodge}

The metric $G$ on $M$ yields a procedure to map, for example, a vector-valued p-form linearly to a covector-valued twisted (n-p)-form, where $n = \mbox{dim}(M)$. This map can be thought of as a vector bundle map $L(\bigwedge_p(T M); TM) \rightarrow L(\tilde{\bigwedge}_{n-p}(T M); T^*M)$ over the identity on $M$, or the resulting linear map given by $\Gamma$. (The operation of $\Gamma$ on vector bundle maps was defined in the second paragraph of subsection \ref{ela_energy_section}.)

We will define linear maps 
\begin{align} \label{Hodge_x}
\star^{\flat}: L(\bigwedge_p(T_X M); T_XM) \rightarrow L(\tilde{\bigwedge}_{n-p}(T_X M); T_X^*M) 
\end{align}
identically for all $X \in M$. For this, recall that the Hodge operator for p-vectors is defined as follows. Given $u \in \bigwedge_p(T_X M)$, we consider the linear map $\bigwedge_{n-p}(T_X M) \rightarrow \bigwedge_n(T_X M)$ defined by $v \mapsto u \wedge v$. Selecting a unit n-vector $\sigma$ for the basis of $\bigwedge_n(T_X M)$, this map may be identified with a linear map $\bigwedge_{n-p}(T_X M) \rightarrow \mathbb{R}$, and the Riesz representation theorem implies the existence of unique $\hat{u} \in \bigwedge_{n-p}(T_X M)$, such that
\begin{align}
u \wedge v = \langle \hat{u}, v \rangle \sigma \ \ \forall v \in \bigwedge_{n-p}(T_X M), \nonumber
\end{align}
where $\langle \cdot , \cdot \rangle$ is the inner product on $\bigwedge_{n-p}(T_X M)$ implied by $G$ (see for example \cite{Kovanen,Flanders}). The twisted (n-p)-vector represented by $(\hat{u}, \sigma)$ (or $(-\hat{u}, -\sigma)$) is uniquely determined by $u$, defining the operator $\star: \bigwedge_p(T_X M) \rightarrow \tilde{\bigwedge}_{n-p}(T_X M)$. Also, the (n-p)-vector $\hat{u}$ is uniquely determined by the twisted p-vector represented by $(u, \sigma)$, defining the operator $\star: \tilde{\bigwedge}_{n-p}(T_X M) \rightarrow \bigwedge_p(T_X M)$. Now, we may define the Hodge operator for p-covectors, and further, for elements of $L(\bigwedge_p(T_X M); T_XM)$. For instance, the operator $\star^{\flat}$ in \eqref{Hodge_x} is defined by its operation on any $\eta \in L(\bigwedge_p(T_X M); T_XM)$ as
\begin{align} \label{Hodge_def}
(\star^{\flat} \eta)(\tilde{u})(v) = G(\eta(\star \tilde{u}), v) 
\end{align}
for all $\tilde{u} \in \tilde{\bigwedge}_{n-p}(T_X M)$ and $v \in T_XM$.

The definition \eqref{Hodge_def} has the following consequence. Using a basis $(v_1,\dots,v_n)$ for $T_XM$, and its dual basis $(\alpha^1,\dots,\alpha^n)$ to express any $\eta \in L(\bigwedge_p(T_X M); T_XM)$ as $v_i \otimes \eta_{j_1 < \dots < j_p}^i \alpha^{j_1} \wedge \dots \wedge \alpha^{j_p}$ (summation over repreated indices with $j_1 < \dots < j_p$), we have
\begin{align}
\star^{\flat} \eta =  (v_i)^{\flat} \otimes \star (\eta_{j_1 < \dots < j_p}^i \alpha^{j_1} \wedge \dots \wedge \alpha^{j_p}), \nonumber
\end{align}
where $\star: \bigwedge^p(T_X M) \rightarrow \tilde{\bigwedge}^{n-p}(T_X M)$ is the Hodge operator of p-covectors, and $(v_i)^{\flat} \in T_X^*M$ is the covariant version of $v_i$. By reusing the notation $\langle \cdot, \cdot \rangle$ for the inner product on $L(\bigwedge_p(T_X M); T_XM)$, defined by
\begin{align}
\langle \eta, \beta \rangle = G_{ki} G^{l_1 j_1} \dots G^{l_p j_p} \eta^k_{l_1 < \dots < l_p} \beta^i_{j_1 \dots j_p},
\end{align}
we have
\begin{align}
\eta \dot{\wedge} \star^{\flat} \beta = \langle \eta, \beta \rangle \mbox{Vol},
\end{align}
where $\mbox{Vol} \in \tilde{\bigwedge}^n(T_X M)$ is the volume element implied by $G$. 

The preceding yields the Hodge operator
\begin{align}
\star^{\flat}: \Gamma(L(\bigwedge_p(T M); TM)) \rightarrow \Gamma(L(\tilde{\bigwedge}_{n-p}(T M); T^*M)). \nonumber 
\end{align}
Similar construction yields, for instance, the Hodge operator 
\begin{align}
\star^{\sharp}: \Gamma(L(\tilde{\bigwedge}_p(T M); T^*M)) \rightarrow \Gamma(L(\bigwedge_{n-p}(T M); TM)), \nonumber
\end{align}
whose definition uses the inverse of the metric correspondence $\cdot^{\flat}$.

\subsection{Operations on vector-valued 1-forms} \label{strain_classical}

In the following we will need the trace $\mbox{tr}$, taking vector-valued 1-forms linearly to real-valued functions. For this, linear maps $\mbox{tr}: L(T_XM;T_XM) \rightarrow \mathbb{R}$ are defined identically for all $X \in M$, by setting $\mbox{tr}(\varepsilon) = \varepsilon_i^i$ (sum over $i=1,\dots,n$). The definition is independent of the used basis. The resulting vector bundle map, and the linear map given by $\Gamma$ (see the first paragraph of subsection \ref{local_forces}), will also be denoted as $\mbox{tr}$.

Since a section $\varepsilon$ of $L(TM;TM)$ may be considered as a vector bundle map $\hat{\varepsilon}: TM \rightarrow TM$ over the identity on $M$, we can form the composition $\hat{\varepsilon} \circ \hat{\varepsilon}$, which is again a vector bundle map $TM \rightarrow TM$ over the identity on $M$, and can therefore be considered as a section of $L(TM;TM)$. It will be denoted as $\varepsilon^2$. 

\section{Magnetostatics}

In the present paper we take a restricted view of electromagnetism, allowing only static observers (with respect to the rigid inertial observer $o$ of section \ref{elastostatics}) that sense pure magnetic field. This way relativistic effects are avoided, and we may use the classical space-time model of section \ref{setting}. In preparation to formulate the coupled magneto-elastic problem, we wish to choose for the problem domain a reference manifold $\Omega$ that has the material body $M$ as its submanifold. For this, we need to transfer the magnetic quantities observed on $S$ to the reference manifold $\Omega$. For static formulation, this means taking into account the particular configuration of $M$ which corresponds to the observer's measurement of the magnetic quantities. 

In the following, the formulation will be given directly on a 3-dimensional reference manifold $\Omega$ with boundary $\partial \Omega$, and its relation to measurements will be considered in section \ref{example}, where constitutive laws will be specified. We will emphasize the analogy of magnetostatics to the formulation of elasticity given above. We will be brief and show only the main points, as it is straightforward to fill in the details by using the analogy with the previous section. 

\subsection{Variational formulation} \label{var_magnetic}

Let us model \emph{magnetic induction} $b$ as a 2-form, that is, an element of $\Gamma(\bigwedge^2(T \Omega))$. The absence of magnetic charges is imposed by expressing $b$ as the exterior derivative of $a \in \Gamma(T^*\Omega)$, that is, we require
\begin{align}
b = \xdif a,
\end{align}
where $a$ will be called \emph{magnetic covector potential}. This equation is invariant to changes of allowable observers. For, the change of observer is represented by a diffeomorphism $\chi: \Omega \rightarrow \Omega$, and the exterior derivative is natural with respect to diffeomorphisms, that is $\chi^* \circ \xdif = \xdif \circ \chi^*$. Therefore, denoting as $\hat{b}$ and $\hat{a}$ the transformed magnetic induction and magnetic covector potential, that is, $\hat{b} = (\chi^{-1})^* b$ and $\hat{a} = (\chi^{-1})^* a$, we have $\hat{b} = \xdif \hat{a}$.

To record the virtual work done by electrical energy sources to generate a variation $\delta a \in \Gamma((T^*\Omega))$ of the magnetic covector potential, we introduce the \emph{current density} as a section of $L(T^*\Omega, \tilde{\bigwedge}^3(T \Omega))$. With hindsight, it is denoted as $\tilde{j} \wedge$. (As the notation suggests, there is an isomorphism between $L(T^*\Omega, \tilde{\bigwedge}^3(T \Omega))$ and $\tilde{\bigwedge}^2(T \Omega)$, but this will be required only later.) Also, we introduce a \emph{surface magnetic field intensity} as a section of $L(T^*\partial \Omega, \tilde{\bigwedge}^2(T \partial \Omega))$, denoted as $\tilde{h}_s \wedge$. Let us then introduce \emph{magnetic drive force} $F^m$ as the linear functional on $\Gamma(T^*\Omega)$ defined by
\begin{align} \label{mag_force}
F^m(\delta a) = \int_{\Omega} \tilde{j} \wedge \delta a + \int_{\partial \Omega} \tilde{h}_s \wedge i^*\delta a
\end{align}
for all $\delta a \in \Gamma(T^*\Omega)$, where $i: \partial \Omega \rightarrow \Omega$ is the natural inclusion. The term $\tilde{h}_s \wedge$ may not be given beforehand on some part of $\partial \Omega$, and then it will be defined on this part by using the notion of magnetic field intensity (section \ref{Ampere}). The equation \eqref{mag_force} is invariant to changes of allowable observers.

For a variational formulation of magnetism, we introduce \emph{magnetic energy density} as a fiber bundle morphism $\tilde{\Phi}: \bigwedge^2(T \Omega) \rightarrow \tilde{\bigwedge}^3(T \Omega)$ over the identity on $\Omega$. Therefore, at each point $X \in \Omega$ we have a smooth map $\tilde{\Phi}_X: \bigwedge^2(T_X \Omega) \rightarrow \tilde{\bigwedge}^3(T_X \Omega)$ (quadratic in case of linear constitutive laws). Magnetic energy density allows us to define \emph{magnetic energy} as the map
\begin{align}
W_m: \Gamma(T^* M) \rightarrow \mathbb{R}; a \mapsto \int_{\Omega} \tilde{\Phi}(\xdif a), 
\end{align}
where $\tilde{\Phi}$ is considered as a (non-linear) map $\Gamma\big(\bigwedge^2(T \Omega)\big) \rightarrow \Gamma\big(\tilde{\bigwedge}^3(T \Omega)\big)$, and integration is a map $\Gamma\big(\tilde{\bigwedge}^3(T \Omega)\big) \rightarrow \mathbb{R}$. This expression for magnetic energy is independent of allowable observers. The transformed magnetic energy density under $\chi: \Omega \rightarrow \Omega$ is $\hat{\tilde{\Phi}} = (\chi^{-1})^* \circ \tilde{\Phi} \circ \chi^*$, and the magnetic energy is given by integrating $\tilde{\Phi}(\xdif a) = \chi^* \big(\hat{\tilde{\Phi}}(\xdif \hat{a})\big)$ over $\Omega$. Therefore, the following variational procedure may be carried out in case of any allowable observer.

To consider variations of magnetic energy density that result from variations of magnetic induction, we take the derivatives of the restricted maps $\tilde{\Phi}_X$ corresponding to all $X \in \Omega$. That is, we define, for each magnetic induction value $b$ (section of $\bigwedge^2(T \Omega)$), a section $D \tilde{\Phi}(b)$ of $L\big(\bigwedge^2(T \Omega); \tilde{\bigwedge}^3(T \Omega)\big)$ by
\begin{align}
D \tilde{\Phi}(b)(X) = \xDif(\tilde{\Phi}_X)(b(X))
\end{align}
for all $X \in \Omega$. The section $D \tilde{\Phi}(b)$ can operate on magnetic induction variations (sections of $\bigwedge^2(T \Omega)$) to produce energy density variations (sections of $\tilde{\bigwedge}^3(T \Omega)$).

To state the variational principle for magnetostatics we assume that magnetic energy density $\tilde{\Phi}$ and current density $\tilde{j} \wedge$ are specified on $\Omega$, and that $\tilde{h}_s \wedge$ is specified on some (possibly empty) part of $\partial \Omega$. Further, $i^* a$ is predefined on the part $S_m \subset \partial \Omega$ where $\tilde{h}_s$ is not specified. Then, we take as a basic principle that the virtual work done by external electrical energy sources coincides with the variation of magnetic energy for all admissible magnetic covector potential variations. Accordingly\footnote{\label{Mag_energy_note}Similar consideration applies here as in footnote \ref{Elastic_energy_note}.}, the problem is to find $a \in \Gamma(T^*\Omega)$, with $i^* a$ predefined on $S_m$, such that
\begin{equation} \label{var_princip_m}
\int_{\Omega} D \tilde{\Phi}(\xdif a)(\xdif \delta a) = F^m(\delta a)
\end{equation}
for all $\delta a \in \Gamma(T^*\Omega)$ with $i^* \delta a$ zero on $S_m$. If $F^m$ is completely specified beforehand it must be done such that $F^m(\delta a) = 0$ for all variations $\delta a$ with $\xdif \delta a = 0$.

\subsection{The algebraic structure and constitutive law}

Various numerical solution methods make use of the fact that (electro)magnetism may be modeled by using exterior algebra equipped with the exterior derivative and a constitutive law \cite{Bossavit_CEM}. The exterior product 
\begin{align}
\wedge: \Gamma(\bigwedge^p(T \Omega)) \times \Gamma(\bigwedge^q(T \Omega)) \rightarrow \Gamma(\bigwedge^{p+q}(T \Omega))
\end{align}
may be constructed in an analogous manner to what we did with the product $\dot{\wedge}$ in section \ref{product}. We only state here that $\wedge$ is bilinear, graded anticommutative, and associative, and it makes (letting $n = \mbox{dim}(\Omega)$) the direct sum
\begin{align}
\Gamma(\bigwedge^0(T \Omega)) \oplus \Gamma(\bigwedge^1(T \Omega)) \oplus \cdots \oplus \Gamma(\bigwedge^n(T \Omega))
\end{align}
into a graded algebra. (We have used the notations $\bigwedge^0(T \Omega) = \Omega \times \mathbb{R}$ and $\bigwedge^1(T \Omega) = T^*\Omega$.) When equipped with the exterior derivative, we have an instance of differential graded algebra.

We also have the graded algebra of forms on the boundary $\partial \Omega$. Then, the trace $\mbox{t}: \Gamma(\bigwedge^p(T \Omega)) \rightarrow \Gamma(\bigwedge^p(T \partial \Omega))$ is an algebra homomorphism defined by pulling back forms on $\Omega$ using $i: \partial \Omega \rightarrow \Omega$. (To pull back twisted forms the boundary $\partial \Omega$ must be transverse orientable.) The same symbol $\mbox{t}$ will be used for the trace of forms to (n-1)-dimensional submanifolds of $\Omega$ (material interfaces).

For the constitutive law, we use the magnetic energy density $\tilde{\Phi}$, and define the \emph{magnetic field intensity} $\tilde{h} \in \Gamma(\tilde{\bigwedge}^1(T \Omega))$ by
\begin{align} \label{const_m}
\tilde{h} \wedge = D \tilde{\Phi}(b),
\end{align}
using the isomorphism between $\Gamma(\tilde{\bigwedge}^1(T \Omega))$ and $\Gamma(L(\bigwedge^2(T \Omega); \tilde{\bigwedge}^3(T \Omega))$ provided by the exterior product. This constitutive law is invariant to changes of allowable observers, that is, we have $\hat{\tilde{h}} \wedge = D \hat{\tilde{\Phi}}(\hat{b})$, where $\hat{\tilde{h}} = (\chi^{-1})^* \tilde{h}$. (This can be shown by considering the transformation rule of $\tilde{\Phi}$, applying the chain rule to the pointwise derivative, and making use of the compatibility of pull-back with the exterior product.) Consequently, the derivation of the following Ampere's law is valid in case of any allowable observer.

\subsection{Ampere's law} \label{Ampere}

We use the above algebraic structure to derive Ampere's law and boundary conditions. To allow for discontinuities of $\tilde{h}$ on material interfaces, this 1-form is required to be only piecewise smooth on $\Omega$. Then, equation \eqref{var_princip_m}, together with \eqref{mag_force} and \eqref{const_m}, is equivalent to Ampere's law
\begin{align} \label{amp1}
\xdif \tilde{h} = \tilde{j}
\end{align}
on regularity regions where $\tilde{h}$ is smooth, the interface condition
\begin{align} \label{amp2}
[\mbox{t} \tilde{h}] = 0
\end{align}
setting the discontinuity $[\mbox{t} \tilde{h}]$ of $\mbox{t} \tilde{h}$ to zero on the boundaries of these regularity regions (assuming zero surface current on these interfaces), and the boundary condition
\begin{align} \label{amp3}
- \mbox{t} \tilde{h} = \tilde{h}_s \ \ \mbox{on} \ \ \partial \Omega \setminus S_m.
\end{align}
The verification of this is similar to the calculation performed in subsection \ref{equilibrium_eq} (but easier since now we are dealing with real-valued forms). 

Finally, to define the magnetic drive force of section \ref{var_magnetic}, we set
\begin{align}
\tilde{h}_s = - \mbox{t} \tilde{h} \ \ \mbox{on} \ \ S_m.
\end{align}

\section{Magneto-elasticity} \label{coupling}

In this section, the above described elastic and magnetic problems are coupled by taking both the displacement gradient and the magnetic induction as system state variables. We allow magneto-elastic coupling both through constitutive behavior and through magnetic forces. 

\subsection{Variational formulations}

For a variational formulation of magneto-elasticity, we recall that the Whitney sum $L(T\Omega; T\Omega) \oplus \bigwedge^2(T \Omega)$ of $L(T\Omega; T\Omega)$ and $\bigwedge^2(T \Omega)$ is the vector bundle whose fiber above $X \in \Omega$ is the direct sum $L(T_X\Omega; T_X\Omega) \oplus \bigwedge^2(T_X \Omega)$ (see for example \cite{AMR}). Then, we introduce \emph{magneto-elastic energy density} as a fiber bundle morphism
\begin{align}
\tilde{\Psi}: L(T\Omega; T\Omega) \oplus \bigwedge^2(T \Omega) \rightarrow \tilde{\bigwedge}^3(T \Omega) \nonumber
\end{align}
over the identity on $\Omega$. Accordingly, when restricted to the fiber above $X \in \Omega$, we have a smooth map $\tilde{\Psi}_X: L(T_X\Omega; T_X\Omega) \oplus \bigwedge^2(T_X \Omega) \rightarrow \tilde{\bigwedge}^3(T_X \Omega)$. The 3-form produced when operating by $\tilde{\Psi}$ on section $(\varepsilon, b)$ of $L(T\Omega; T\Omega) \oplus \bigwedge^2(T \Omega)$ will be denoted as $\tilde{\Psi}(\varepsilon, b)$.

To consider variations of energy density that result from displacement gradient variations, we take the partial derivatives of the restricted maps $\tilde{\Psi}_X$, corresponding to all $X \in \Omega$, with respect to the first argument. Accordingly, for given section $(\varepsilon,b)$ of $L(T\Omega; T\Omega) \oplus \bigwedge^2(T \Omega)$, we define a section $D_1 \tilde{\Psi}(\varepsilon,b)$ of $L(L(T\Omega; T\Omega); \tilde{\bigwedge}^3(T \Omega))$ by
\begin{align}
D_1 \tilde{\Psi}(\varepsilon,b)(X) = \xDif\big(\tilde{\Psi}_X(\cdot, b(X))\big)(\varepsilon(X))
\end{align}
for all $X \in \Omega$. Similarly, to consider variations of energy density that result from magnetic induction variations, we take partial derivatives with respect to the second argument. That is, for given section $(\varepsilon,b)$ of $L(T\Omega; T\Omega) \oplus \bigwedge^2(T \Omega)$, we define a section $D_2 \tilde{\Psi}(\varepsilon,b)$ of $L(\bigwedge^2(T \Omega); \tilde{\bigwedge}^3(T \Omega))$ by
\begin{align}
D_2 \tilde{\Psi}(\varepsilon,b)(X) = \xDif\big(\tilde{\Psi}_X(\varepsilon(X),\cdot)\big)(b(X)),
\end{align}
for all $X \in \Omega$.

Next, we describe two different models for magneto-elasticity, having slightly different ranges of applicability.

\subsubsection{Coupling through magneto-elastic energy and magnetic forces} \label{mag_ela1}

The model described here assumes not only small displacement gradients, but also small overall deformations, such that the stress-free reference configuration may be assumed to correspond with the observer's measurement of the magnetic quantities. (Really, the stress-free reference configuration correspond to zero magnetic field, and the observer's measurement of magnetic quantities takes place in the deformed equilibrium configuration, where magnetic forces are balanced by inner stresses in the material.)

In the given model, the modeler directly specifies a rule that gives magnetic stresses from the state variables of the coupled system. This will result in magnetic forces on the right hand side of the equilibrium equation. This means that the modeler should not include in the magneto-elastic energy density $\tilde{\Psi}$ a particular mechanism by which magnetic field is affected by displacement gradient, namely, the mechanism that identifies the magnetic measurements to the deformed equilibrium configuration. For, this particular mechanism is already taken into account on the right hand side of the equilibrium equation by the prescribed magnetic forces. Consequently, as the particular mechanism becomes excluded from the constitutive law, this construct assumes not only small displacement gradients in the material body, but also small overall deformations. For example, this rules out situations with ``long'' material bodies, which may deform significantly despite of small strains in the material body. 

For the coupling through magnetic forces, we proceed in two steps. First, we assume a fiber bundle morphism
\begin{align}
\tilde{\mathcal{S}}: L(T\Omega; T \Omega) \oplus \bigwedge^2(T \Omega) \rightarrow L(\tilde{\bigwedge}_2(T \Omega); T^*\Omega)
\end{align}
over the identity on $\Omega$, which will be called \emph{magnetic stress mapping}. In the simplest cases the maps $\tilde{\mathcal{S}}_X: L(T_X\Omega; T_X\Omega) \oplus \bigwedge^2(T_X \Omega) \rightarrow L(\tilde{\bigwedge}_2(T_X \Omega); T_X^*\Omega)$ will be quadratic in $b_X$, and independent of $\varepsilon_X$ (section \ref{example}). When operating on a section $(\varepsilon, b)$ of $L(T\Omega; T \Omega) \oplus \bigwedge^2(T \Omega)$ this will give a covector-valued 2-form, denoted as $\tilde{\mathcal{S}}(\varepsilon, b)$. The above magnetic stress mapping allows the kind of magnetic stresses used typically in engineering, see for example \cite{Fonteyn}.

As the next step, we replace the force functional of section \ref{force} by a map $F: \Gamma(L(T\Omega; T \Omega) \oplus \bigwedge^2(T \Omega)) \times \Gamma(TM) \rightarrow \mathbb{R}$, defined such that
\begin{align} \label{mag_force_fun}
F((\varepsilon,b), \delta \nu) = \int_M (\tilde{f} + \xdif_{\nabla} \tilde{\mathcal{S}}(\varepsilon,b)) \dot{\wedge} \delta \nu + \int_{\partial M} (\tilde{\tau} + [\mbox{t} \tilde{\mathcal{S}}(\varepsilon,b)]) \dot{\wedge} i^* \delta \nu
\end{align}
for all $((\varepsilon,b), \delta \nu) \in \Gamma(L(T\Omega; T \Omega) \oplus \bigwedge^2(T \Omega)) \times \Gamma(TM)$. Here, the term $[\mbox{t} \tilde{\mathcal{S}}(\varepsilon,b)]$ is the trace of $\tilde{\mathcal{S}}(\varepsilon,b)$ taken from the outside of $M$ minus the trace taken from the inside. If $\tilde{\tau}$ is not given beforehand on some part of $\partial M$, it will be defined on this part by using stress.

The coupled problem may now be stated as follows. Let us be given magneto-elastic energy density $\tilde{\Psi}$, magnetic stress mapping $\tilde{\mathcal{S}}$, the mechanical (as opposed to magnetic) body force $\tilde{f} \dot{\wedge}$ on $M$, and the mechanical surface force $\tilde{\tau} \dot{\wedge}$ on some part $\partial M \setminus S_e$ of $\partial M$, the current density $\tilde{j} \wedge$ on $\Omega$, and the surface magnetic field intensity $\tilde{h}_s \wedge$ on some part $\partial \Omega \setminus S_m$ of $\partial \Omega$. Then, using the inclusions $i_M: \partial M \rightarrow M$ and $i_{\Omega}: \partial \Omega \rightarrow \Omega$, the problem is to find $\nu \in \Gamma(T M)$ and $a \in \Gamma(T^*\Omega)$, with $i_M^* \nu$ predefined on $S_e$ and $i_{\Omega}^* a$ predefined on $S_m$, such that
\begin{equation} \label{var_princip_coup1_e}
\int_M D_1 \tilde{\Psi}(\nabla \nu, \xdif a)(\nabla \delta \nu) = F((\nabla \nu, \xdif a), \delta \nu)
\end{equation}
for all $\delta \nu \in \Gamma(T M)$ with $i_M^* \delta \nu$ zero on $S_e$, and
\begin{equation} \label{var_princip_coup1_m}
\int_{\Omega} D_2 \tilde{\Psi}(\nabla \nu, \xdif a)(\xdif \delta a) = F^m(\delta a)
\end{equation}
for all $\delta a \in \Gamma(T^*\Omega)$ with $i_{\Omega}^* \delta a$ zero on $S_m$. 

We may now use the differential graded algebra of exterior forms, and the calculus of vector- and covector-valued forms developed in section \ref{elastostatics}, to derive the equilibrium equations of the coupled system. First, for strain $\varepsilon$ and magnetic induction $b$,  we define $\tilde{\sigma} \in \Gamma(L(\tilde{\bigwedge}_2(T \Omega); T^*\Omega))$ by
\begin{align} \label{const_coup1}
\tilde{\sigma} \dot{\wedge} = D_1 \tilde{\Psi}(\varepsilon, b),
\end{align}
and $\tilde{h} \in \Gamma(\tilde{\bigwedge}^1(T \Omega))$ by
\begin{align} \label{const_coup2}
\tilde{h} \wedge = D_2 \tilde{\Psi}(\varepsilon, b).
\end{align}
For convenience, we use the notation $\tilde{\sigma}_{mag}$ for $\tilde{\mathcal{S}}(\varepsilon,b)$. Then, by using \eqref{mag_force_fun} and \eqref{const_coup1} in \eqref{var_princip_coup1_e}, we get the equilibrium equations
\begin{align} 
- \xdif_{\nabla} \tilde{\sigma} & = \tilde{f} + \xdif_{\nabla} \tilde{\sigma}_{mag} \ \ \mbox{on } M, \label{equi_coup1} \\
\mbox{t} \tilde{\sigma} & = \tilde{\tau} + [\mbox{t} \tilde{\sigma}_{mag}] \ \ \mbox{on } \partial M \setminus S_e.  \label{equi_coup2}
\end{align}
By using \eqref{mag_force} and \eqref{const_coup2} in \eqref{var_princip_coup1_m}, we get the familiar \eqref{amp1}-\eqref{amp3}.

Finally, to define the map $F$, we set
\begin{align}
\tilde{\tau} = \mbox{t} \tilde{\sigma} - [\mbox{t} \tilde{\sigma}_{mag}] \ \ \mbox{on } S_e,
\end{align}
and to define the magnetic drive force, we set
\begin{align}
\tilde{h}_s = - \mbox{t} \tilde{h} \ \ \mbox{on} \ \ S_m.
\end{align}

This construction allows for magnetostrictive behavior, in which case there is genuine constitutive coupling between magnetism and elasticity \cite{Bozo}. Here, it is further possible to decompose the energy density into two parts. That is, one can define ``elastic'' energy density $\tilde{\Psi}_e$ and ``magnetic'' energy density $\tilde{\Psi}_m$ as fiber bundle morphisms $L(T\Omega; T\Omega) \oplus \bigwedge^2(T \Omega) \rightarrow \tilde{\bigwedge}^3(T \Omega)$, such that
\begin{align} \label{energy_decomp}
\tilde{\Psi} = \tilde{\Psi}_e + \tilde{\Psi}_m,
\end{align}   
where the sum is defined fiberwise (using the vector space structure of the fibers of $\tilde{\bigwedge}^3(T \Omega)$). For example, we may define $\tilde{\Psi}_e$ independently of $b$, by $\tilde{\Psi}_e(\varepsilon,b) = \tilde{\Psi}(\varepsilon,0)$, and then $\tilde{\Psi}_m$ is defined by $\tilde{\Psi}_m = \tilde{\Psi} - \tilde{\Psi}_e$. In case of magnetostrictive behavior, this $\tilde{\Psi}_m$ depends on $\varepsilon$, and then the use of \eqref{energy_decomp} in the constitutive law \eqref{const_coup1} yields a decomposition 
\begin{align}
\tilde{\sigma} = \tilde{\sigma}_e + \tilde{\sigma}_m,
\end{align}
where the ``elastic'' stress $\tilde{\sigma}_e$ is defined independently of $b$ by $\tilde{\sigma}_e \dot{\wedge} = D_1 \tilde{\Psi}_e(\varepsilon, b)$, and
the ``magnetostrictive'' stress $\tilde{\sigma}_m$ is defined by $\tilde{\sigma}_m \dot{\wedge} = D_1 \tilde{\Psi}_m(\varepsilon, b)$. This results in a ``magnetostrictive'' force in the equilibrium equations for $\tilde{\sigma}_e$. (This is in addition to the magnetic force term in \eqref{equi_coup1} and \eqref{equi_coup2}.)

When magnetostrictive behavior is allowed the above decomposition of energy density is not unique. We may instead define $\tilde{\Psi}_m$ independently of $\varepsilon$, by $\tilde{\Psi}_m(\varepsilon,b) = \tilde{\Psi}(0,b)$, and then $\tilde{\Psi}_e$ is defined by $\tilde{\Psi}_e = \tilde{\Psi} - \tilde{\Psi}_m$. Because now $\tilde{\Psi}_m$ is independent of $\varepsilon$, no ``magnetostrictive'' forces appear. (Instead, one can decompose $\tilde{h}$ to get an additional ``current density'' term to the Ampere's law written for the ``magnetic'' part of $\tilde{h}$.)

\subsubsection{Coupling through magneto-elastic energy} \label{mag_ela2}

In the model described here, all mechanisms by which magnetic field is affected by displacement gradient are included in the magneto-elastic energy density $\tilde{\Psi}$. The model allows taking into account that the measurement of magnetic quantities takes place in the deformed equilibrium configuration, and not in the stress-free reference configuration. Accordingly, it may be used in case of small displacement gradients in the material body, and large overall deformations. This construction, when restricted from the beginning to Euclidean geometry, is considered in \cite{Bossavit}.

Because the coupling is realized totally by using magneto-elastic energy density, the notion of magnetic force is not introduced in the first place. Consequently, we use the force functional $F: \Gamma(TM) \rightarrow \mathbb{R}$ of subsection \ref{force}.

The coupled problem may be stated as follows. Let us be given the magneto-elastic energy density $\tilde{\Psi}$, the mechanical (as opposed to magnetic) body force $\tilde{f} \dot{\wedge}$ on $\Omega$ (zero on $\Omega \setminus M$), and the mechanical surface force $\tilde{\tau} \dot{\wedge}$ on some part $\partial M \setminus S_e$ of $\partial M$, the current density $\tilde{j} \wedge$ on $\Omega$, and the surface magnetic field intensity $\tilde{h}_s \wedge$ on some part $\partial \Omega \setminus S_m$ of $\partial \Omega$. Then, we find $\nu \in \Gamma(T \Omega)$ and $a \in \Gamma(T^*\Omega)$, with $i_M^* \nu$ predefined on $S_e$ and $i_{\Omega}^* a$ predefined on $S_m$, such that
\begin{equation} \label{var_princip_coup2_e}
\int_{\Omega} D_1 \tilde{\Psi}(\nabla \nu, \xdif a)(\nabla \delta \nu) = F(\delta \nu)
\end{equation}
for all $\delta \nu \in \Gamma(T \Omega)$ with $i_M^* \delta \nu$ zero on $S_e$, and
\begin{equation} \label{var_princip_coup2_m}
\int_{\Omega} D_2 \tilde{\Psi}(\nabla \nu, \xdif a)(\xdif \delta a) = F^m(\delta a)
\end{equation}
for all $\delta a \in \Gamma(T^*\Omega)$ with $i_{\Omega}^* \delta a$ zero on $S_m$. We emphasize that \eqref{var_princip_coup2_e} and \eqref{var_princip_coup2_m} are not yet sufficient to solve for $(\nu, a)$, and, as an additional requirement, $\nu$ is extended smoothly to $\Omega \setminus M$, such that $i^* \nu$ vanishes on $\partial \Omega$. The energy density $\tilde{\Psi}$ must be so specified\footnote{This seems to lack rigorous justification, although in \cite{Bossavit_07} it is shown that the derivative of magnetic energy with respect configuration (and thus displacement) is independent of the particular extension of the configuration to the ``electrically passive'' region.}  that $a$ and $\nu$ on $M$, and total system energy, will not depend on the particular extension of $\nu$. (For an example where the extension of $\nu$ is performed according to Laplace equation, see \cite{Kovanen2}.)

The equilibrium equations may be derived as follows. First, for strain $\varepsilon$ and magnetic induction $b$, we define $\tilde{\sigma} \in \Gamma(L(\tilde{\bigwedge}_2(T \Omega); T^*\Omega))$ by
\begin{align} \label{const_coup2_1}
\tilde{\sigma} \dot{\wedge} = D_1 \tilde{\Psi}(\varepsilon, b),
\end{align}
and $\tilde{h} \in \Gamma(\tilde{\bigwedge}^1(T \Omega))$ by
\begin{align} \label{const_coup2_2}
\tilde{h} \wedge = D_2 \tilde{\Psi}(\varepsilon, b).
\end{align}
We assume $\tilde{\sigma}$ only piecewise smooth to allow for ``magnetic surface forces''.
From \eqref{var_princip_coup2_e}, using \eqref{global_force_def} and \eqref{const_coup2_1}, we may now infer the equations
\begin{align} 
- \xdif_{\nabla} \tilde{\sigma} & = \tilde{f} \ \ \mbox{on } M, \label{equi_coup1_2} \\
[\mbox{t} \tilde{\sigma}] & = \tilde{\tau} \ \ \mbox{on } \partial M \setminus S_e,  \label{equi_coup2_2}
\end{align}
and from \eqref{var_princip_coup2_m}, using \eqref{mag_force} and \eqref{const_coup2_2}, we get the familiar \eqref{amp1}-\eqref{amp3}. In \eqref{equi_coup2} the term $[\mbox{t} \tilde{\sigma}]$ is the trace of $\tilde{\sigma}$ taken from the inside of $M$ minus the trace taken from the outside.

Finally, to define the force functional, we set
\begin{align}
\tilde{\tau} = [\mbox{t} \tilde{\sigma}] \ \ \mbox{on } S_e,
\end{align}
and to define the magnetic drive force, we set
\begin{align}
\tilde{h}_s = - \mbox{t} \tilde{h} \ \ \mbox{on} \ \ S_m.
\end{align}

\section{Examples} \label{example}

\subsection{Coupling through magnetic forces}

In the present example, magneto-elastic coupling is due merely to magnetic forces (no magnetostriction). Also, we consider linear isotropic material behavior. The magneto-elastic energy density is expressed as the sum
\begin{align} \label{energy_sum}
\tilde{\Psi} = \tilde{\Psi}_e + \tilde{\Psi}_m,
\end{align}
where the elastic and magnetic energy densities $\tilde{\Psi}_e$ and $\tilde{\Psi}_m$ are defined as follows. The term $\tilde{\Psi}_e$ is defined independently of $b$, such that
\begin{equation} \label{energy_dens_ela}
\tilde{\Psi}_e(\varepsilon,b) = \frac{1}{2} \Big(\lambda \mbox{tr}(\varepsilon)^2 + \mu \big(\mbox{tr}(\varepsilon^2) + \langle \varepsilon, \varepsilon \rangle \big)\Big) \mbox{Vol},
\end{equation}
where $\lambda$ and $\mu$ are the Lam\'e constants, $\langle \cdot, \cdot \rangle$ is the inner product of vector-valued 1-forms implied by the metric $G$, and $\mbox{Vol}$ is the volume form implied by $G$. The term $\tilde{\Psi}_m$ is defined independently of $\varepsilon$, such that 
\begin{align} \label{energy_dens_mag}
\tilde{\Psi}_m (\varepsilon,b) = \frac{1}{2} r \star b \wedge b,
\end{align}
where $r$ is the reluctivity (piecewise smooth function, taken here as element of the multiplicative ring of the module of 1-forms), and $\star$ is the Hodge operator of 2-forms implied by $G$. This specifies the observer's measurement of magnetic quantities to the stress-free reference configuration.

For the elastic constitutive law, we have $\tilde{\sigma} \dot{\wedge} = D_1 \tilde{\Psi}(\varepsilon, b) = D_1 \tilde{\Psi}_e(\varepsilon, b)$, resulting in
\begin{equation} \label{const_ela}
\tilde{\sigma} = \lambda \mbox{tr}(\varepsilon) \star^{\flat} I + 2 \mu \star^{\flat} \mbox{sym}(\varepsilon),
\end{equation}
where $I$ is the identity vector-valued 1-form. One can verify that this constitutive law gives zero variations $\tilde{\sigma} \dot{\wedge} \nabla \delta \nu$ of elastic energy density in case $\nabla \delta \nu$ is antisymmetric, that is, in case $\mathcal{L}_{\delta \nu} g$ vanishes. For the magnetic constitutive law, we have $\tilde{h} \wedge = D_2 \tilde{\Psi}(\varepsilon,b) = D_2 \tilde{\Psi}_m(\varepsilon,b)$, resulting in
\begin{align}
\tilde{h} = r \star b.
\end{align}

To specify the magnetic stress mapping $\tilde{\mathcal{S}}$, we use the interior product of differential forms by vector fields. For p-form $\alpha$ and vector field $u$, their interior product $\mbox{i}_u \alpha$ is the (p-1)-form defined such that $\mbox{i}_u \alpha(v_1,\dots,v_{p-1}) = \alpha(u,v_1,\dots,v_{p-1})$ for all vector fields $v_1,\dots,v_{p-1}$ (see for example \cite{Frankel,AMR}). Now the mapping $\tilde{\mathcal{S}}$ is defined independently of $\varepsilon$, such that
\begin{align} \label{Max_tensor}
\tilde{\mathcal{S}}(\varepsilon,b) \dot{\wedge} v = \frac{1}{2} r \star b \wedge \mbox{i}_v b + \frac{1}{2} \mbox{i}_v (r \star b) \wedge  b
\end{align}
for all vector fields $v$. With Euclidean metric, this is the classical Maxwell's stress tensor, as measured in the reference configuration.

\subsection{Coupling through magneto-elastic energy}

We give an example with Euclidean geometry. As in the previous example, we assume negligible magnetostriction. This means that magneto-elastic coupling will be due merely to change in magnetic energy that results from change in geometry (change in shape of the material body \cite{Bossavit}).

As in the previous example, we restrict to linear isotropic material behavior. The magneto-elastic energy density is the sum of elastic and magnetic energy densities $\tilde{\Psi}_e$ and $\tilde{\Psi}_m$. The term $\tilde{\Psi}_e$ is defined by \eqref{energy_dens_ela} as above, and $\tilde{\Psi}_m$ is defined such that
\begin{align} \label{energy_dens_mag_2}
\tilde{\Psi}_m (\varepsilon,b) = \frac{1}{2} r \hat{\star} b \wedge b,
\end{align}
where $\hat{\star}$ is the Hodge operator of 2-forms implied by a metric $\hat{G}$ that depends on $G$ and $\varepsilon$ as follows.
The metric $\hat{G}$ is defined by
\begin{align} \label{pull_back_metric_gen}
\hat{G}_{ij} & = G_{ik} \big(\delta_j^k + \varepsilon_j^k + G_{jl}\varepsilon_m^l G^{km} + G_{lm} \varepsilon_n^m G^{kn} \varepsilon_j^l\big), 
\end{align}
where $\varepsilon_j^k$ etc. denote the components of $\varepsilon$, and $\delta_j^k$ are the components of the identity vector-valued 1-form. 

The above energy density specifies the observer's measurement of magnetic quantities to the deformed equilibrium configuration. In the Euclidean geometry, the displacement $\nu$ yields the change of configuration, and then the metric $\hat{G}$ of \eqref{pull_back_metric_gen}, expressed in terms of $\varepsilon = \nabla \nu$, is the pull-back of $G$ by the change of configuration. (For details, see \cite{Marsden}, pp. 57-60.) Observe that the metric $\hat{G}$ is defined also in $\Omega \setminus M$, because displacement is extended smoothly to this region.

We may define $\tilde{\sigma}_{mag}$ by $\tilde{\sigma}_{mag} \dot{\wedge} = D_1\tilde{\Psi}_m(\varepsilon, b)$, and then we will get the equilibrium equations
\begin{align}
- \xdif_{\nabla} \tilde{\sigma} & = \tilde{f} + \xdif_{\nabla} \tilde{\sigma}_{mag} \ \ \mbox{on } M, \label{equi_example1} \\
\mbox{t} \tilde{\sigma} & = \tilde{\tau} + [\mbox{t} \tilde{\sigma}_{mag}] \ \ \mbox{on } \partial M \setminus S_e,  \label{equi_example2}
\end{align}
just as equations \eqref{equi_coup1} and \eqref{equi_coup2} in section \eqref{mag_ela1}, but therein the dependence of $\tilde{\sigma}_{mag}$ on $\varepsilon$ originates from material behavior.

Finally, having assumed small displacement gradients, we may neglect the term in \eqref{pull_back_metric_gen} which is quadratic in $\varepsilon$ (at least in the material body $M$). Using this approximation for the metric, and performing the derivation $D_1\tilde{\Psi}_m(\varepsilon, b)$, yields magnetic stress $\tilde{\sigma}_{mag}$ which coincides with the familiar Maxwell's stress tensor, as measured in the deformed equilibrium configuration.

But if magnetic stresses are measured in the equilibrium configuration, so should the magnetic forces, and for this, we should use the magnetic forces $\xdif_{\hat{\nabla}} \tilde{\sigma}_{mag}$, where $\hat{\nabla}$ is the Levi-Civita connection of $\hat{G}$. However, this does not arise from the above formalism. It seems that the setup of small-strain elasticity and the setup of taking account the change of geometry in magnetics does not perfectly fit together. A consistent formulation could be achieved by considering geometrically non-linear elasticity, where the balance of forces and stresses is enforced in the deformed equilibrium configuration.

\section{Conclusion}

The formulation of small-strain elasticity by using some of the recent developments of differential geometry strengthens similarities between small-strain elasticity and magnetism. The given perspective point to the use in small-strain magneto-elasticity the same discretization methods that have already proved powerful in electromagnetic field computation. To make such a program systematic, we suggest searching for algebraic category where a class of magnetic, elastic, and magneto-elastic problems may be modeled, and which constitutes suitable domain for discretizations.

\appendix

\section{The isomorphism of stresses} \label{isomorphism}

Let us verify that the map $L\big(\bigwedge_2(T_XM); T_X^*M\big) \rightarrow L\big(L(T_XM;T_XM); \bigwedge^3(T_XM)\big)$ with operation $\omega \mapsto \omega \dot{\wedge}$,
where $\omega \dot{\wedge}$ is defined by 
\begin{align} \label{def}
(\omega \dot{\wedge} e) (u,v,w) = \omega(u,v)(e(w)) + \omega(w,u)(e(v)) + \omega(v,w)(e(u)) \nonumber
\end{align}
for all $e \in L(T_XM;T_XM)$ and $u,v,w \in T_XM$, is linear isomorphism.

For linearity, let us select arbitrary $\alpha, \beta \in \mathbb{R}$, $\omega_1, \omega_2 \in L\big(\bigwedge_2(T_XM); T_X^*M\big)$, $e \in L(T_XM;T_XM)$, and $u,v,w \in T_XM$. We have
\begin{align}
\big((\alpha \omega_1 + \beta \omega_2) \dot{\wedge} e\big)(u,v,w) & = (\alpha \omega_1 + \beta \omega_2)(u,v)(e(w)) + (\alpha \omega_1 + \beta \omega_2)(w,u)(e(v)) \nonumber \\
& \ \ + (\alpha \omega_1 + \beta \omega_2)(v,w)(e(u)) \nonumber \\
& = \big(\alpha (\omega_1(u,v)) + \beta (\omega_2(u,v))\big)(e(w)) + \dots \nonumber \\
& = \Big(\alpha \big(\omega_1(u,v)(e(w))\big) + \beta \big(\omega_2(u,v)(e(w))\big)\Big) + \dots \nonumber \\
& = \alpha(\omega_1 \dot{\wedge} e)(u,v,w) + \beta(\omega_2 \dot{\wedge} e)(u,v,w) \nonumber \\
& = \big(\alpha(\omega_1 \dot{\wedge} e) + \beta(\omega_2 \dot{\wedge} e)\big)(u,v,w) \nonumber
\end{align}
Therefore, we have the equality of the 3-covectors
\begin{align}
(\alpha \omega_1 + \beta \omega_2) \dot{\wedge} e = \alpha(\omega_1 \dot{\wedge} e) + \beta(\omega_2 \dot{\wedge} e), \nonumber
\end{align}
and thus the equality 
\begin{align}
(\alpha \omega_1 + \beta \omega_2) \dot{\wedge} = \alpha(\omega_1 \dot{\wedge}) + \beta(\omega_2 \dot{\wedge}) \nonumber
\end{align}
for all $\alpha, \beta \in \mathbb{R}$ and $\omega_1, \omega_2 \in L\big(\bigwedge_2(T_XM); T_X^*M\big)$.

For bijectivity, let us first select a basis $(v_1,v_2,v_3)$ for the tangent space $T_XM$, and denote as $(\alpha^1,\alpha^2,\alpha^3)$ the dual basis of $T_X^*M$. Then, the tensors $\alpha^i \otimes (\alpha^j \wedge \alpha^k)$, where $i,j,k = 1,2,3$ with $j < k$, constitute a basis for $L\big(\bigwedge_2(T_XM); T_X^*M\big)$. Further, since the tensors $v_j \otimes \alpha^i$, where $i,j = 1,2,3$, constitute a basis for the vector space $L(T_XM, T_XM)$, the dual basis $(e_i^j)_{i,j=1,2,3}$ of $L(L(T_XM, T_XM); \mathbb{R})$ is defined by
\begin{align}
e_i^j (v_k \otimes \alpha^l) = \left\{ \begin{array}{ll}
1 & \mbox{for } k=j \mbox{ and } l=i\\
0 & \mbox{otherwise.} \end{array} \right. \nonumber
\end{align}
Then, a basis for $L\big(L(T_XM;T_XM); \bigwedge^3(T_XM)\big)$ is given by $(\alpha^1 \wedge \alpha^2 \wedge \alpha^3) \otimes e_i^j$, where $i,j = 1,2,3$. In these bases the map $\omega \mapsto \omega \dot{\wedge}$ is given by (using summation over repeated indices with $j < k$)
\begin{align}
(\omega \dot{\wedge})_i^l = \epsilon^{jkl} \omega_{ij<k}, \nonumber
\end{align}
where $\epsilon^{jkl}$ is the permutation symbol. (This symbol takes the value $+1$ or $-1$ when the indices $i,j,k$ form an even or odd permutation of $1,2,3$, respectively, otherwise it takes the value zero.) Thus, the linear map becomes represented by diagonal $9 \times 9$ matrix, with $\pm 1$ on the diagonal, which is a non-singular matrix.


\end{document}